\documentstyle[aps,twocolumn,epsfig]{revtex}
\begin{document}

\draft
\title{             COSMOLOGY IN A TEST TUBE: \\
        THEORY OF DOMAIN WALLS FORMATION IN BINARY FLUIDS }

\author{Jacek Dziarmaga$^{1,2}$
        \thanks{e-mail address: dziarmaga@t6-serv.lanl.gov}
        and Mariusz Sadzikowski$^3$}
\address{ 1) Los Alamos National Laboratory, Theoretical
             Astrophysics T-6, MS B288, Los Alamos, NM87545, USA\\
          2) Institute of Physics, Jagiellonian University,
             Krak\'ow, Poland\\
          3) Institute of Nuclear Physics, Radzikowskiego 152,
             31-342 Krak\'ow, Poland}
\date{November 5, 1999}
\maketitle
\tighten


\begin{abstract}

{\bf
Formation of domain walls during a rapid phase transition in a quasi one
dimensional Cahn-Hiliard equation describing binary fluids in a thin tube
is studied. Density of kinks scales like a sixth root of quench rate for
equal concentrations and like a square root of quench rate for unequal
concentrations of fluids. For a slow inhomogeneous transition the density
is linear in velocity of temperature front. This paper is first
theoretical study of topological defects formation in a system with
conserved order parameter.
}

\end{abstract}

  It has been pointed out some time ago that topological defects which
formed during subsequent symmetry breaking phase transitions can provide
seeds for structure formation in the early universe \cite{book}. Kibble
\cite{kibble} gave a detailed theory of defect formation in I order phase
transitions which proceed by bubble nucleation. Bubbles are born with
random orientation of order parameter; when they coalesce they can give
rise to a nontrivial topological winding number. This prediction was
verified in relatively simple beautiful experiments in liquid crystals
\cite{lq}. Disclinations were observed and theoretical relation between
bubble density and disclination density was verified. All this at nearly
room temperature and with at most the aid of an optical microscope.

  A scenario for II order transitions was put forward by Zurek
\cite{zurek}. He observed that the order parameter goes out of equilibrium
some time before the critical point is crossed. In this way a unique
finite correlation length $\hat{\xi}$ proportional to the forth root of
the quench rate is frozen-in. $\hat{\xi}$ is the scale which determines
density of topological defects after the transition. Experimental
verification is not as clear as for I order transitions. Experiments in
superfluid $^4He$ were done \cite{4He} just to be falsified later on.
There are spectacular experiments in superfluid $^3He$ \cite{3He}. Density
of detected vortices is consistent with the order of magnitude Zurek's
prediction. However the tricky detection is somewhat indirect, there is no
way to control the quench rate and verify the scaling of $\hat{\xi}$ with
the rate, finally the precise origin of detected vortices was also
recently put in question \cite{akv}. There is urgent need for an
experiment as simple as that in liquid crystals.

  A good candidate for such an experiment are binary fluids described by
Cahn-Hiliard plus Navier-Stokes equations. Order parameter is conserved
and proportional to the difference in fluid concentrations. An example are
aniline and cycloheksan. They do not mix for temperatures less than
$T_c=30.9^oC$. The transition is second order with a standard
Ginzburg-Landau free energy. The fluids differ in optical density and thus
domain walls between them can be detected by optical means. A rapid quench
should result in a quench rate dependent density of domain walls. The best
experimental approach is to quench the fluids in a thin tube, which makes
the system effectively one dimensional. In one dimension the quench-made
domain walls cannot be eradicated by mutual annihilation thanks to order
parameter conservation - they are a permanent record of the nonequilibrium
transition. The transition in one dimension is, strictly speaking, a
crossover; the correlation length does not diverge at $T_c$. This
deviation from a mean-field is important only in a very narrow critical
regime which is irrelevant for our nonequlibrium quenches. In fact we do
not see any deviation from a mean-field in our numerical simulations.

  Dynamics of binary fluid separation in one spatial dimension is
described by model H dynamics which consists of the Cahn-Hiliard (CH)
equation for conservative real order parameter $\phi$

\begin{equation}\label{CH1}
\dot\phi=\left[\; -\epsilon(t,x)\phi+\phi^3-\phi'' \;\right]''+\xi' \;\;,
\end{equation}
where $\dot{}=\partial_t$ and $'=\partial_x$. CH equation is usually
supplemented by Navier-Stokes equation but in one dimension the
incompressibility condition makes this perturbation trivial. We allow for
variation of the symmetry breaking parameter $\epsilon$
(temperature/pressure) both in space and in time. $\xi$ is assumed to be a
white Gaussian noise with nonvanishing cumulants

\begin{eqnarray}\label{noise}
&& \langle \xi(t,x) \rangle=0 \;\;,  \nonumber\\
&& \langle \xi(t_1,x_1) \xi(t_2,x_2) \rangle=
   2T\;\delta(t_1-t_2)\;\delta(x_1-x_2) \;\;.
\end{eqnarray}

{\bf TRANSITION WITH EQUAL CONCENTRATIONS. } To begin with let us consider
a uniform linear phase transition with

\begin{equation}\label{linear}
\epsilon(t,x)=\frac{t}{\tau} \;\;,
\end{equation}
with $\phi=0$ on average at the initial $t=-\infty$, which is preserved by
CH evolution. Any uniform transition close to the critical point at
$\epsilon=0$ can be described by such a linearized $\epsilon$.

 For $t<0$ the system is in a symmetric phase; $\phi$ is subject to small
fluctuations around $0$. For this stage of the quench and also for the
onset of spinodal instability, just after $\epsilon$ crossed $0$,
Eq.(\ref{CH1}) can be linearized in $\phi$,

\begin{equation}\label{l}
\dot\phi=\left[\; - \frac{t}{\tau} + 
                  3 \langle \phi^2 \rangle \;\right]\; \phi'' -
         \phi'''' + \xi'   \;\;.
\end{equation}
The mean-field $\langle \phi^2 \rangle$ is kept just to control validity
of the linearization. We solve Eq.(\ref{l}) by neglecting the mean-field
term and performing a Fourier transformation

\begin{equation}\label{Fourier}
\phi(t,x)=\int_{-\infty}^{+\infty} dk\; e^{ikx} \phi(t,k) \;\;.
\end{equation}
Eq.(\ref{l}) can be solved for any $k$ with a help of its Green function
and the correlations (\ref{noise}). The power spectrum $P(t,k)$ of the
fluctuations is

\begin{eqnarray}\label{P}
&& \langle \phi^\star (t,k) \phi(t,p) \rangle \equiv
   P(t,k)\delta(k-p) \;\;, \nonumber \\
&& P(t,k)=\;\frac{T\sqrt{\tau}|k|}{\pi}\;
            e^{\frac{k^2(t-\tau k^2)^2}{\tau}}\;
            \int_{-\infty}^{ \frac{|k|(t-\tau k^2)}{\sqrt{\tau}} }
            ds\; e^{-s^2} \;.
\end{eqnarray}
The fluctuations, measured by $\langle\phi^2\rangle=\int dk\;P(t,k)$, are
small for $t<0$. At some $\hat{t}>0$

\begin{equation}
\langle \phi^2 \rangle\;\equiv\;
\int_{-\infty}^{+\infty}dk\; P(\hat{t},k)\;
\approx \;
\frac{\hat{t}}{\tau} \;\equiv\; \hat\epsilon
\end{equation}
and the linearized approximation involved in solving Eq.(\ref{l}) breaks
down. Assuming that $t/\tau^{2/3}>0$ is sufficiently large the error
function integral in (\ref{P}) is constant for small $k^2$ (but it is
still essential to suppress the divergence for large $k^2$). The remaining
exponent is peaked at $k^2\approx t/3\tau$ with a maximum of
$\exp(4t^3/27\tau^2)$. This maximum begins to blow up at
$\hat{t}\approx\tau^{2/3}$. It is around this time that $\langle \phi^2
\rangle$ passes through $t/\tau$. This is the moment when kinks of width
given by the corresponding healing length of $\tau^{1/6}$ begin to form.
Fluctuations with $|k|>\hat{k}=\tau^{-1/6}$ are irrelevant for kink
formation at $\hat{t}$. Density of kinks which are going to form can be
identified with the density of zeros of $\phi(\hat{t},x)$ smoothed over
$|k|>\hat{k}$ which is, according to a formula from \cite{mazenko},

\begin{equation}\label{mazenko}
n=\frac{\pi}{2}
\sqrt{ \frac{ \int_{-\hat{k}}^{+\hat{k}}dk\;k^2\;P(\hat{t},k) }
            { \int_{-\hat{k}}^{+\hat{k}}dk\;P(\hat{t},k) }} \;.
\end{equation}
Introducing an integration variable $k/\hat{k}$ one can see 
that any $\tau$-dependence can be factorized in front of the
integrals so that 

\begin{equation}\label{n}
n \;\sim\; \frac{1}{\tau^{1/6}} 
\end{equation}
for any $\tau$. Results from numerical simulations consistent with this
prediction are shown in Fig.1. In the limit of adiabatic transition
$\tau\rightarrow\infty$ the system stays close to the critical point for a
time long enough to order at long distance. For a fast quench substantial
amount of initial disorder is frozen into the ordered phase in a form of
kinks. In one dimension, because of $\phi$-conservation, these domain
walls have no freedom to diffuse around and mutually annihilate. There is
no "phase ordering kinetics" to erase this trace of disorder. The kinks
are permanent record of the phase transition.

{\bf TRANSITION WITH UNEQUAL CONCENTRATIONS. } The two fluids may differ
by average concentration. In that case the conserved average $\phi$ is
$M\neq 0$. We take $M>0$ for definiteness.  An uniform $\phi=M$
configuration is stable against small perturbations if $\epsilon<3 M^2$
when $M$ is outside the interval between the two inflection points of the
double-well potential. For $\epsilon<M^2\;$ $\phi=M$ is not a minimum of
the effective potential but it cannot decay because of
$\phi$-conservation; $\phi=M$ is bigger than the positive minimum at
$\sqrt{\epsilon}$ for $\epsilon >0$ or the minimum at $0$ for $\epsilon
<0$.

  At $\epsilon=M^2\;$ $\phi=M$ coincides with the minimum of the
potential. When $M^2<\epsilon<3M^2\;$ $\phi=M$ is stable against small
perturbations but again it is not a minimum of the double well potential.
This time, however, $\phi$-conservation does not forbid its decay to the
minima at $\pm\sqrt{\epsilon}$. The decay can proceed by thermal
nucleation of antikink-kink pairs (AKP) provided that $T$ is large as
compared to a barrier. To estimate minimal $T$ necessary for nucleation
let us introduce $\epsilon=3M^2+\varepsilon$ with $\varepsilon<0$ and
expand $\phi=M+\tilde{\phi}$. The effective potential can be approximated
by $\frac12|\varepsilon|\tilde{\phi}^2$. Fluctuations around $M$ are
$\langle \tilde{\phi}^2 \rangle\sim T/\sqrt{|\varepsilon|}$. They can
result in AKP nucleation if they reach beyond the positive inflection
point at $\sqrt{(3M^2-|\varepsilon|)/3}$ or in other words
$(M-\sqrt{(3M^2-|\varepsilon|)/3})^2\approx\langle \tilde{\phi}^2
\rangle$. To first order in $|\varepsilon|/M^2$ nucleation takes place at
$|\varepsilon| < (TM^2)^{2/5}=\varepsilon_n$. Nucleation time is $\sim
\varepsilon_n^{-2}$. If the transition proceeds at a finite rate,
$\varepsilon=t/\tau$, there may be not enough time for thermal nucleation.
Nucleation can only happen if at $\varepsilon=-\varepsilon_n$ the
nucleation time is shorter than the time left till $\varepsilon$ reaches
$0$. This condition is satisfied if

\begin{equation}\label{nucl}
\tau(TM^2)^{6/5} \; \gg \; 1 \;\;.
\end{equation}
If the transition is fast enough or $T$ is sufficiently small, when the
opposite condition holds, no AKP are nucleated for $M^2<\epsilon<3M^2$;
$\phi$ remains fluctuating around $M$ until $\epsilon$ crosses $3M^2$ and
spinodal decomposition due to instability of $\phi=M$ begins. This case
can be analyzed as follows. As a first step we define
$\epsilon(t)=3M^2+t/\tau$ and expand $\phi=M+\tilde{\phi}$. Equation
(\ref{CH1}) when linearized in $\tilde{\phi}$ gives Eq.(\ref{l}) but with
$\phi$ replaced by $\tilde{\phi}$. By the same token as for equal
concentrations, at $\hat{t}=\tau^{2/3}$ fluctuations blow up with momentum
peaked at $\pm\hat{k}\sim\pm\tau^{-1/6}$. In contrast to the $M=0$ case,
the growth of this instability is halted very quickly much before any
kinks are formed. To see this let us pick a wave
$\tilde{\phi}=a(t)\cos(\hat{k}x)$. At around $\hat{t}=\tau^{2/3}$ maxima
of this wave enter the area beyond the inflection point of the actual
double-well potential, $\phi>\sqrt{(3M^2+\tau^{-1/3})/3}$ - their growth
is slowed down. At the same time minima get further and further into the
unstable regime between the inflection points so it seems that their
growth rate should be accelerated. If it were so they would quickly hop to
the neighborhood of the negative inflection point and the initial
$\cos(\hat{k}x)$ would distort into a periodic array of kinks-antikinks
with density $\sim\hat{k}$. It cannot be so because it would obviously
violate $\phi$-conservation. The growth of $a(t)$ stops as soon as the
maxima cross the inflection point. At this stage the cosine wave is still
an almost negligible fluctuation around the large $M$, especially for
large $\tau$. However the $\hat{k}$ scale is frozen-in as a record of the
quench. That is what remains of the spinodal stage. AKP nucleation begins.

  The problem with the $\cos(\hat{k}x)$ lies with its perfect periodicity
which cannot be broken. However the power spectrum although peaked at
$\sim\pm\hat{k}$ is also spread around them by $\sim\hat{k}$. This means
in particular that some minima of the "cosine" are more negative than the
other. There is competition between them. Each of them wants to hop to the
negative inflection point but the number of winners is limited by
$\phi$-conservation - the average $\phi$ must remain $M$.

  A fraction of winners can be estimated in a following way. At the end
of the spinodal stage we have a landscape of correlated $\tilde{\phi}$
domains of size $\hat{k}^{-1}$. A fraction $q$ of negative domains will
win. Those negative domains who win will end at the negative inflection
point $\phi=-\sqrt{M^2+1/3\tau^{1/3}}$. All the positive domains and the
negative loosers will end at $\phi=+\sqrt{M^2+1/3\tau^{1/3}}$. To keep
$\phi=M$ on average, $q$ must satisfy

\begin{equation}\label{balance}
M=\frac{[-q+(1-q)]+1}{2}\sqrt{M^2+1/3\tau^{1/3}} \;\;,
\end{equation}
where $[-q+(1-q)]$ comes from positive and $1$ from negative domains. For
$M^2\tau^{1/3} \gg 1$ we obtain $q\approx 1/6M^2\tau^{1/3}$ so the kink
density should scale as
 
\begin{equation}
n \;\sim\; q\hat{k}^{-1} \;\sim\; \frac{1}{M^2\tau^{1/2}} \;\;\;
\mbox{for} \;\;
M^2\tau^{1/3} \gg 1 \;.
\end{equation}
This $1/2$ exponent is $3$ times bigger than the $1/6$ in Eq.(\ref{n}). It
should be much easier measurable in experiment. Numerical simulations are
consistent with this prediction as shown in Fig.1. The process of
spinodal decomposition followed by nucleation is shown in Fig.2.

{\bf INHOMOGENEOUS TRANSITION. } An uniform phase transition
(\ref{linear}) may be a good first approximation in some cases but in real
life we often have to face the fact that it is not perfectly homogeneous.
To gain some insight let us begin with the temperature front of the form

\begin{equation}\label{front}
\epsilon(t,x)= \left\{
\begin{array}{cc}
+\epsilon_{-} \;, & x<vt \\
-\epsilon_{+} \;, & vt<x 
\end{array}
\right.
\end{equation}
in the limit of very slow $v$. We will argue {\it a posteriori} that the
sharp step is a good approximation of any generic front for $v \rightarrow
0$.

 We solve the problem of kink generation behind the moving front by
perturbative expansion around $v=0$. At $v=0$ there is a static
$\phi$-front, $\phi(x)=H(x)$ - a step in $\phi$ at $x \approx 0$
interpolating between $-\sqrt{\epsilon_{-}}$ at $x=-\infty$ and $0$ at
$x=+\infty$. Its width is given by $\approx
\epsilon_{-}^{-1/2}+\epsilon_{+}^{-1/2}$.

 Let us now switch on small $v>0$. The $\epsilon$-front (\ref{front}) is
slowly moving on. If $\phi$ were not conserved, the $\phi$-front would
follow moving in step with $\epsilon$-front and leaving no kinks behind
\cite{inhom}. For our conserved $\phi$ it is not possible, kinks must
inevitably appear. To see in some detail how it happens let us substitute
$\phi(t,x)=H(x-vt)+\psi(t,x)$ with $\psi=O(v)$ to Eq.(\ref{CH1}) and keep
only $O(v)$ terms. We are interested in length scales large as compared to
the width of the step $H(x)$ and that of the $\epsilon$-front. That is why
we keep only up to the second $x$-derivative which is responsible for
diffusion. Far from $x \approx vt$ we obtain

\begin{equation}\label{dpsi}
\dot\psi(t,y)=\epsilon_{+}\psi''(t,y)+v\psi'(t,y) +
              v\sqrt{\frac{\epsilon_{+}}{2}}\;\theta(t)\delta(y) \;,
\end{equation}
where $y=x-vt$. We take into account a $\delta$-like source term at $y=0$
which is a long-wavelength approximation to $vH'$. We also set
$\epsilon_{-}=\epsilon_{+}/2$ for simplicity. The source term is switched
on at $t=0$ when the $\epsilon$-front starts to move, hence the Heaviside
function. The solution of Eq.(\ref{dpsi}) is straightforward,

\begin{equation}\label{psi}
\psi(t,y)=
v\sqrt{\frac{\epsilon_{+}}{2}}
\int_{0}^{t}dt'\;
\frac{ e^{-\frac{ [y+v(t-t')]^2 }{ 4\epsilon_{+}(t-t') }} }
     { \sqrt{4\pi\epsilon_{+}(t-t')} } \;\;.
\end{equation}
The source term produces $\psi$ at $y=0$ at constant rate. It spreads
around by diffusion but at the same time it is carried to negative $y$
with velocity $-v$. For $y>0$ the penetration by diffusion dominates at
first but at $v^2t^2\sim\epsilon_{+}t$ the two processes balance one
another and $\psi(t,y>0)$ saturates. From this time on all the $\psi$ is
carried directly from the source to $y<0$ with velocity $-v$. This
effectively means that the $\phi$-front halted, while the $\epsilon$-front
keeps moving on. A supercooled phase with a slightly positive $\phi$ is
growing in between them with velocity $v$. When its width exceeds
$\sqrt{2/\epsilon_{+}}\;$, $\phi$ decays towards positive ground state.
From this time on we have a negative $\phi$-step moving together with
$\epsilon$-step and the whole story repeats itself at spatial intervals of
$\epsilon_{+}/v$. Density of kinks is

\begin{equation}
n \; \sim \; \frac{v}{\epsilon_{+}} \;\;
\mbox{for} \;\; v \rightarrow 0 \;\;.
\end{equation}
It should be stressed that the whole process is deterministic, kinks are
made at regular intervals. Noise is required to begin the process; it also
adds some irregularity on top of the regular pattern.

  Note that for small $v$ the relevant length scale is $\epsilon_{+}/v$.
For small enough $v$, it far exceeds the $\epsilon$-front width and the
width of $H(x)$. This justifies the sharp step in Eq.(\ref{front}) and the
long-wavelength approximations involved in our derivation of
Eq.(\ref{dpsi}).

  Let us now turn to the opposite large-$v$ limit where we anticipate the
transition to be effectively homogeneous. Any generic $\epsilon(t,x)$ can
be linearized around $\epsilon=0$,

\begin{equation}\label{inhlin}
\epsilon(t,x)=\frac{vt-x}{v\tau}\equiv\alpha(vt-x) \;\;.
\end{equation}
At any fixed $x$ the transition proceeds at the rate of $1/\tau$ just like
in Eq.(\ref{linear}). If it were homogeneous it would enhance the momentum
$\hat{k}=\tau^{-1/6}$. For

\begin{equation}
v \gg \alpha^5 \;\;\mbox{or}\;\;
v \gg \tau^{-5/6}  \;\;
\end{equation}
this momentum scale is much bigger than the slope $\alpha$ and the
relevant field fluctuations do not feel the inhomogeneity.  This is where
the transition is effectively homogeneous and Eq.(\ref{n}) applies.

{\bf CONCLUSION. } We studied dynamics of domain walls formation during a
quench in effectively one dimensional binary fluids. We believe that our
predictions can be verified by experiments in thin test tubes. In fact
first steps in this direction has been already made \cite{exp}. One can
test the $1/6$ scaling for equal concentrations, which is the most direct
analogue of Kibble-Zurek scenario for second order transitions. A more
intriguing case are unequal concentrations where we also get scaling with
the quench rate in a range of parameters. Its $1/2$ exponent should be
much easier to measure than the $1/6$. Finally, unlike for nonconserved
order parameter \cite{inhom}, density of kinks is linear in front velocity
for a slow inhomogeneous quench. We believe that binary fluids are a
unique opportunity of an almost bare eye detection of topological defects.
We also believe that one can not only detect topological defects but also
proceed and measure scaling of their density with quench rate. In this
paper we lay theoretical foundations for the experimental work in progress
\cite{exp}.

{\bf Acknowledgements.} We would like to thank Henryk Arod\'z, Maciej
Nowak and Wojciech \.Zurek for useful comments. M.S. was partially
supported by KBN grant 2P03B 086 14.

\begin{figure}
\centerline{\epsfxsize=6 cm \epsfbox{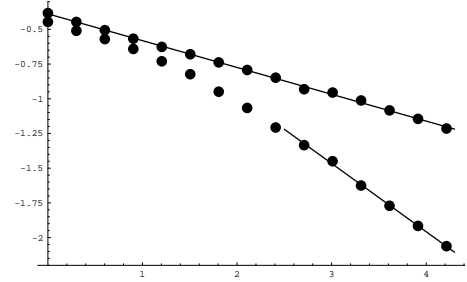}}

\caption{ $\log(n)$ as a function of $\log(\tau)$ for $M=0$ (the top plot)
and for $M=2$ (the bottom plot) according to numerical simulations. For
$M=0$ the slope is $0.18\pm 0.01$ as compared to the theoretical
$1/6\approx0.17$. For $M=2$ the slope saturates for $\log(\tau)>2$ at
$0.48\pm0.05$ as compared to the theoretical $0.50$. At low $\tau$ the
$M=0,2$ results tend to be the same. Vertical size of a point is its
statistical error. Simulations were done at $T=10^{-5}$ on a $\Delta x=1$,
$\Delta t=0.01$ lattice of size $1024$ with periodic boundary conditions.
$\epsilon(t)$ was swept from $3M^2-10\tau^{-1/3}$ to $3M^2+10\tau^{-1/3}$.
Kinks were counted at final time. Density $n$ is an average over many
runs.}

\end{figure}

\begin{figure}
\centerline{\epsfxsize=6 cm \epsfbox{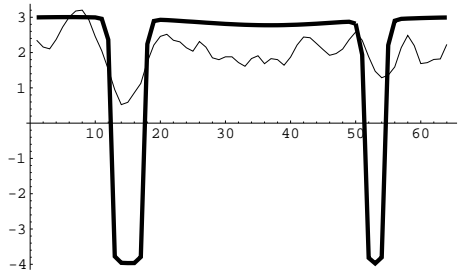}}

\caption{ Two snapshots of $\phi$ as a function of $x$ for $M=2,\tau=128,
T=10^{-5}$ taken from numerical simulations. The thin line is $\phi$ at
$t=0.8\hat{t}$ when spinodal decomposition begins. The amplitude of
fluctuations around $M=2$ is magnified 100 times. The thick line is $\phi$
at $t=1.6\hat{t}$ when kinks are already well defined.}

\end{figure}

\end{document}